# Borderline magnetism in $Sr_4Ru_3O_{10}$: Impact of dilute La and Ca doping on itinerant ferromagnetism and metamagnetism

S. Chikara, V. Durairaj, W.H. Song[†], Y.P. Sun[†], X.N. Lin, A. Douglass and G. Cao*

Department of Physics and Astronomy, University of Kentucky

Lexington, KY 40506

P. Schlottmann

Department of Physics and National High Magnetic Field Laboratory,

Florida State University, Tallahassee, FL 32306

(Dated: January 22, 2006)

Abstract

An investigation of La and Ca doped $Sr_4Ru_3O_{10}$, featuring a coexistence of interlayer ferromagnetism and intralayer metamagnetism, is presented. La doping readily changes magnetism between ferromagnetism and metamagnetism by tuning the density of states. It also results in *different* Curie temperatures for the c-axis and the basal plane, highlighting a rare spin-orbit coupling with the crystal field states. In contrast, Ca doping enhances the c-axis ferromagnetism and the magnetic anisotropy. La doping also induces a dimensional crossover in the interlayer transport whereas Ca doping exhibits a tunneling magnetoresistance and an extraordinary $T^{3/2}$-dependence of the resisitivity. The drastic changes caused by the dilute doping demonstrate a rare borderline magnetism that is delicately linked to the interplay of the density of states and spin-orbit coupling.

# I. Introduction

Understanding itinerant ferromagnetism and metamagnetism is a longstanding challenge in Condensed Matter Physics [1,2]. According to the Stoner model [3], the condition for spontaneous ferromagnetism requires that the Coulomb exchange interaction, U, is strong and, in addition, the density of states at the Fermi surface, $g(E_F)$, is large, so that $Ug(E_F) \geq 1$, which is known as the Stoner criterion. If $Ug(E_F)$ is large but not sufficiently close to 1 (i.e., $Ug(E_F)<1$), enhanced paramagnetism characterized by a large and temperature-dependent magnetic susceptibility is expected. The detailed properties of the Stoner enhanced $\chi(T)$ are determined by the energy dependence of $g(E)$ in the vicinity of the Fermi level. Peaks of $g(E)$ are often related to Van Hove singularities and intimately coupled to magnetism and phonons (lattice deformations). Hence, the field-induced itinerant metamagnetism [2,4-6] observed in several materials such as $Sr_3Ru_2O_7$ [7,8], $Y(Co_{1-x}Al_x)_2$ [5] and other Co compounds [6] is believed to be induced by a nearby Stoner instability. Recent studies on correlated metals such as MnSi [9] and $Sr_3Ru_2O_7$ [7,8,10] reveal phenomena consistent with quantum criticality due to the onset of itinerant ferromagnetism and the critical end-point of a first-order metamagnetic transition, respectively. The essence of this physics has been captured by a simple model [11] invoking a minimum of $g(E)$ (MnSi) and a two-dimensional Van Hove singularity for the ruthenate. Clearly, itinerant ferromagnetism and metamagnetism sensitively depend on U and $g(E_F)$, and are not expected to coexist. But $Sr_4Ru_3O_{10}$ under doping defiantly shows the coexistence of both. It is this coexistence that suggests new physics.

$Sr_4Ru_3O_{10}$ belongs to the layered ruthenate series, $(Ca, Sr)_{n+1}Ru_nO_{3n+1}$ (n=number of Ru-O layers/unit cell). Rich with novel physical phenomena rarely found in other materials, these materials share as a central feature the extended 4d-electron orbitals, which lead to comparable and thus competing energies for crystalline fields (CEF), Hund's rule interactions, spin-orbit coupling, p-d hybridization and electron-lattice coupling. The deformations and relative orientations of corner-shared $RuO_6$ octahedra crucially determine the CEF level splitting and the band structure, and hence the nature of the ground state. As a result, the physical properties are highly dimensionality (or n) dependent and susceptible to perturbations such as the application of magnetic fields, pressure and slight changes in chemical compositions (electron-lattice coupling). These characteristics are illustrated in $Ca_{n+1}Ru_nO_{3n+1}$ and $Sr_{n+1}Ru_nO_{3n+1}$ (n=1, 2, 3, ∞): The former are on the verge of a metal-insulator transition and prone to antiferromagnetism that changes with n, whereas the latter are metallic, and evolve from paramagnetism (n=1, 2) to a ferromagnetic state (n=∞) with increasing n [7,8,10,12-31].

Situated between n=2 and n=∞, the triple-layered $Sr_4Ru_3O_{10}$ (n=3) displays complex phenomena ranging from tunneling magnetoresistance, low frequency quantum oscillations [25-28] to switching behavior [29]. The most intriguing feature, however, is borderline magnetism: While *along the c-axis* (perpendicular to the layers), $Sr_4Ru_3O_{10}$ shows ferromagnetism ($Ug(E_F) \geq 1$) with a saturation moment $M_S$ of 1.13 $\mu_B$/Ru and a Curie temperature $T_C$ at 105 K followed by increased spin polarization below $T_M$=50 K, it features for the field in *the ab-plane* a sharp peak in the magnetization at $T_M$=50 K and a first-order metamagnetic transition [26], a situation strikingly similar to Stoner enhancement ($Ug(E_F)<1$) responsible for the enhanced paramagnetism and itinerant metamagnetism [4-8,11]. The coexistence of the interlayer ferromagnetism and the intralayer metamagnetism, i.e. the anisotropy in the field response, is not expected from simple theoretical arguments [1,2,6], and has then to arise from the two-dimensional Van Hove singularity (logarithmical divergence) close to the Fermi level [11] in conjunction with the coupling of the spins to the crystalline field orbital states and the lattice. $Sr_4Ru_3O_{10}$ (n=3) is therefore a unique system that is delicately positioned on the borderline separating its closest neighbors, the ferromagnet $SrRuO_3$ (n=∞) [20] and the paramagnet $Sr_3Ru_2O_7$ (n=2) [7,8], and provides a rare opportunity to study itinerant ferromagnetism and metamagnetism by slightly tuning $g(E_F)$ through band filling (electron doping) and bandwidth control (structural alteration).

In this paper we report results of our study on $Sr_4Ru_3O_{10}$ with $Sr^{2+}$ being replaced by small amounts of $La^{3+}$ and $Ca^{2+}$ ions. Probing magnetism with these substitutions offers following advantages: A concentration x of $La^{3+}$ dopes the system with x electrons on the Ru sites, altering $g(E_F)$ and the exchange splitting $\Delta$. In addition, the $La^{3+}$ and $Ca^{2+}$ ions are significantly smaller than the $Sr^{2+}$ ion (the ionic radii: $r_{La}$=1.03 Å and $r_{Ca}$=1.00 Å, compared to $r_{Sr}$=1.18 Å); hence, low concentration doping enhances the buckling of the $RuO_6$ octahedra, varying the exchange interaction or bandwidth while preserving the crystal structure. Because of the similarity of the ionic size, the impact of the La and Ca doping on the structural distortions is expected to be similar. Therefore, studying and comparing responses to La and Ca doping not only reveals new phenomena, but also differentiates the effect of electron doping and the structural distortion on the itinerant magnetism. Indeed, properties of $(Sr_{1-x}La_x)_4Ru_3O_{10}$ and $(Sr_{1-x}Ca_x)_4Ru_3O_{10}$ with 0≤x≤0.13 [32] vary widely and drastically. Most significantly, La doping effectively reduces $g(E_F)$, $\Delta$ and $M_S$, leading to an evolution from ferromagnetism to metamagnetism along the c-axis but a reverse development within the basal plane. It also results in *different* Curie temperatures for the c-axis and the basal plane that bring to light an unusual interplay of the spin-orbit coupling with the CEF states. In sharp contrast, Ca doping enhances the c-axis spontaneous ferromagnetism, but drastically weakens the basal plane magnetization. In terms of transport properties, La doping induces a dimensional crossover in the interlayer transport at high temperatures and Ca doping results in a large tunneling magnetoresistance at x=0.02 and an unusual $T^{3/2}$-power law for the resistivity at x=0.13 below $T_C$, suggesting non-Fermi-liquid behavior [9]. The large array of novel phenomena presented illustrates the rare borderline magnetism that is critically determined by the interplay of the density of the states and spin-orbit coupling with the crystal field states of the $RuO_6$-octahedra.

**II. Results and discussion**

Fig.1 shows the temperature dependence of magnetization M for $(Sr_{1-x}La_x)_4Ru_3O_{10}$ for (a) the c-axis and (b) the ab-plane, and for $(Sr_{1-x}Ca_x)_4Ru_3O_{10}$ for (c) the c-axis and (d) the ab-plane. This figure contrasts the impact of the La and Ca doping on M(T). As seen in Fig.1a, the magnetization along the c-axis, $M_c$, displays a gradual evolution from the ferromagnetism to paramagnetism with increasing x, as manifested by the rapid decrease of $T_C$, indicated by the vertical arrows. Upon cooling the transition at $T_M$ (denoted with arrowheads), which for x=0 marks the increase in $M_c$ at 50 K, develops into a sharp downturn for x=0.05, 0.08 and 0.11, and eventually into a peak at x=0.13, signaling the entry into the paramagnetic state. In contrast, within the ab-plane ferromagnetism occurs upon La doping at $T_M$ where $M_c$ has its maximum (Fig.1b). In Fig.

1b the arrowheads indicate the onset of ferromagnetism. It is striking that $M_{ab}$ hardly shows an anomaly at $T_C$, where $M_c$ has such a pronounced dependence (compare Fig.1a and b). Evidently, the La-doping causes a strong anisotropy in M favoring ferromagnetism along the c-axis for $T_M<T<T_C$, but within the ab-plane for $T<T_M$. On the other hand, Ca doping preserves the temperature dependence of $M_c$ and visibly increases $T_C$ and $T_M$, as well as for $M_{ab}$ for sufficiently small x, but entirely changes $M_{ab}$ for x=0.13 (Fig.1c and 1d). The arrows and arrowheads highlight the key features of the temperature dependence of M.

Fig.2 shows the isothermal magnetization M(B) at T=2 K for $(Sr_{1-x}La_x)_4Ru_3O_{10}$ for (a) the c-axis and (b) the ab-plane, and for $(Sr_{1-x}Ca_x)_4Ru_3O_{10}$ for (c) the c-axis and (d) the ab-plane. For x=0, the $M_c(B)$ is readily saturated with increasing B at 0.2 T, yielding a $M_S$ of 1.13 $\mu_B$/Ru, i.e. more than a half of the 2 $\mu_B$/Ru expected for an S=1 system and comparable to that of $SrRuO_3$ [20]. Metamagnetic behavior develops with increasing x and becomes well-defined in $M_c(B)$ for x≥0.08 as seen in Fig.2a. This is consistent with the enhanced paramagnetism for $T<T_M$ in $M_c(T)$ shown in Fig.1a. On the other hand, $M_{ab}(B)$ shows a first-order metamagnetic transition at $B_c$ (=2.5 T at 2 K) for x=0. This metamagnetic transition essentially disappears for x>0.05, where $M_{ab}(T)$ shows ferromagnetic behavior (Fig.1b). The impact of Ca doping is different, since the ferromagnetism $M_c(B)$ strengthens (Fig.2c), but $M_{ab}(B)$ shows a higher $B_c$ (=3.5 T at 2 K) for x=0.02. The metamagnetism in the ab-plane then disappears at x=0.13 and is replaced by a nearly linear field dependence as shown in Fig.2d, indicative of the vanishing ferromagnetism in the basal plane. Figs.3a and 3b highlight the major impacts of La and Ca doping and their differences: the La doping effectively reduces $T_C$, $T_M$ (dashed lines) and $M_S$ (solid lines). However, the Ca doping enhances ferromagnetism along the c-axis, but weakens $M_{ab}$. The arrows representing the spins in the three layers schematically describe the effects of the La and Ca doping on the spin configuration.

Shown in Fig. 4 is the temperature dependence of resistivity, $\rho$, at B=0 for $(Sr_{1-x}La_x)_4Ru_3O_{10}$ for (a) the c-axis and (b) the basal plane, and for $(Sr_{1-x}Ca_x)_4Ru_3O_{10}$ for (c) the c-axis and (d) the basal plane. For x=0, the c-axis resistivity, $\rho_c$, exhibits anomalies corresponding to $T_C$ and $T_M$, and precipitously drops by an order of magnitude from $T_M$(=50 K) to 2 K due to the reduction of spin-scattering as the spins become strongly polarized below $T_M$ [26]. This drop in $\rho_c$ at low T disappears upon La doping as a result of the strong reduction of spin polarization below $T_M$ (see Fig.1a). The increase in residual resistivity $\rho_o$ can be attributed to an enhancement of the elastic scattering rate $\tau^{-1}$ either due to increased spin-flip scattering and/or to disorder caused by the doping. In either case the contributions to $\tau^{-1}$ are essentially temperature-independent. Remarkably, for T>$T_M$ $\rho_c$ decreases by as much as a factor of 2 with x, but, on the other hand, $\rho_{ab}$ increases significantly with x. This behavior suggests enhanced interlayer hopping but weakened intralayer transport due to doping. It is likely that the La and Ca impurities break the symmetry and give rise to a stronger overlap of the $d_{xz}$ and $d_{yz}$ orbitals and hence to a larger conductivity along the c-axis but a reduced one in the ab-plane due to scattering. In addition, a dimensional crossover is facilitated by substituting the smaller ions that shorten the separation between the two-dimensional layers, a situation possibly similar to the temperature-driven crossover in the interlayer transport in layered materials such as $NaCo_2O4$ and $Sr_2RuO_4$ [33].

The temperature-dependence of $\rho_c$ can be associated with changes in the quasiparticle effective mass $m_{eff}$. With the exception of x=0.05, the Fermi liquid behavior survives up T<17 K for the La doped samples as both $\rho_c$ and $\rho_{ab}$ follow the dependence $\rho = \rho_o + AT^2$, where $A \sim m_{eff}^2$. For x=0, $A_c = 1.04 \times 10^{-5}$ $\Omega$ cm/K$^2$ and $A_{ab} = 3.4 \times 10^{-7}$ $\Omega$ cm/K$^2$, i.e. the $A_c/A_{ab}$(=31) ratio is unusually large, suggesting a strongly anisotropic Fermi surface or $m_{eff}$. $A_c/A_{ab}$ is drastically reduced to 3.9 for x=0.08 and 1.4 for x=0.11 of La doping. This decrease is primarily due to the drop in $A_c$(=2.5×10$^{-7}$ $\Omega$ cm/K$^2$ for x=0.11) because $A_{ab}$(=1.8×10$^{-7}$ $\Omega$ cm/K$^2$ for x=0.11) is only slightly smaller. The smaller $A_c$ implies a smaller $m_{eff}$, therefore larger electron mobility for the interlayer transport.

It needs to be pointed out that the Fermi-liquid behavior is conspicuously violated for x=0.05 of La doping and x=0.13 of Ca doping. First, $\rho_{ab}$ for x=0.05 La doping is exceptionally larger than that for other x. Second, both $\rho_{ab}$ and $\rho_c$ below 17 K obey a $T^{5/3}$-power law as shown (for $\rho_{ab}$) in the inset in Fig.4b. Marginal Fermi-liquid models [1, 34] predict power laws of the resistivity as a function of T with non-integer and even non-universal exponents at low T. The $T^{5/3}$ power law is anticipated [1] when small angle electron scattering dominates the electronic transport, but is rarely observed in a ferromagnetic state far below $T_C$. This $T^{5/3}$-behavior is intrinsic and unlikely to be due to disorder because the Fermi-liquid behavior is recovered when x increases as discussed above. More surprisingly, for Ca doping, both $\rho_{ab}$ and $\rho_c$ for x=0.13 show a $T^{3/2}$-dependence for 3<T<46 K (see the inset in Fig.4c). The $T^{3/2}$-power law, which remains when B is applied, marks the breakdown of the Fermi-liquid properties. Such behavior, which is also observed in other itinerant ferromagnets such as MnSi at high pressure, is believed to be associated with effects of diffusive motion of the electrons caused by the interactions between the itinerant electrons and critically damped magnons [9].

Shown in Fig.5 is the magnetoresistivity ratio, $\Delta\rho_c(B)/\rho_c(0)$ with $\Delta\rho_c(B) = \rho_c(B) - \rho_c(0)$, for Ca doping at x=0.02 as a function of B applied within the basal plane. It features a sharp drop at $B_c$ and reaches a value as large as 40% in the vicinity of and below $T_M$. The large reduction in $\rho_c$ for $B_{\|ab}>B_c$ implies large ferromagnetic fluctuations in a state without ferromagnetic long-range order immediately above the transition. In contrast, $\rho_c$ for x=0.05 La doping shows much smaller $\Delta\rho_c(B)/\rho_c(0)$, suggesting that scattering is much less spin-dependent as a result of the electron doping.

**III. Conclusions**

This work illustrates that the rare borderline magnetism in $Sr_4Ru_3O_{10}$ is highly sensitive to $g(E_F)$ that is critically linked to band filling and structural distortions and that metamagnetism is an immediate neighbor of ferromagnetism. The results indicate that the magnetism determined by $g(E_F)$ in $Sr_4Ru_3O_{10}$ seems to be more susceptible to band filling than to a structural distortion. The different $T_C$ for the c-axis and the basal plane underscore a rare spin-orbit coupling with the crystal field states of the octehedra $RuO_6$. On the other hand, the structural distortion caused by Ca doping enhances the c-axis ferromagnetism but weakens the basal plane magnetism, causing larger magnetic anisotropy. While the transport is intimately coupled to the magnetism, the largely reduced $\rho_c$ for $T>T_M$ signifies a strengthened overlap of $d_{xz}/d_{yz}$ orbitals and a dimensional crossover. The $T^{3/2}$-depedence of the resistivity provides evidence for a breakdown of the Fermi liquid model. All unusual behavior clearly results from the borderline magnetism that calls for new paradigms for studying the itinerant magnetism.

**Acknowledgements**: This work was supported by NSF grant DMR-0240813 and DOE grant DE-FG02-98ER45707

[†]Permanent address: Institute of Solid State Physics, Chinese Academy of Sciences, Hefei 230031, Anhui, P.R. China.

*Corresponding author. cao@uky.edu

**Figure Captions:**

**Fig.1**. The magnetization as a function of temperature at B=0.01 T for $(Sr_{1-x}La_x)_4Ru_3O_{10}$ for the field (a) along the c-axis and (b) in the ab-plane, and for $(Sr_{1-x}Ca_x)_4Ru_3O_{10}$ (c) along the c-axis and (d) in the ab-plane. Note that c-axis $T_C$ is indicated by arrows whereas c-axis $T_M$ and ab-plane $T_C$ are indicated by arrowheads and that c-axis $T_C$ in (a) corresponds to no clear anomalies in the ab-plane in (b).

**Fig.2**. The isothermal magnetization M at T=2 K for $(Sr_{1-x}La_x)_4Ru_3O_{10}$ for the field (a) along the c-axis and (b) in the ab-plane, and for $(Sr_{1-x}Ca_x)_4Ru_3O_{10}$ for (c) along the c-axis and (d) in the ab-plane.

**Fig.3**. Dependence of $T_C$, $T_M$, (dashed lines) $M_S$ (solid lines) on x for (a) La doping and (b) Ca doping. The arrows indicate spins in the triple layers and schematically describe the effects of La and Ca doping on the spin configuration.

**Fig.4**. The resistivity, $\rho$, as a function of temperature for $(Sr_{1-x}La_x)_4Ru_3O_{10}$ for the field (a) along the c-axis and (b) in the basal plane, and for $(Sr_{1-x}Ca_x)_4Ru_3O_{10}$ for (c) along the c-axis and (d) in the basal plane. Inset in panel (b): $\rho_{ab}$ vs. $T^{5/3}$ for La doping at x=0.05. Inset in panel (c): $\rho_{ab}$ vs. $T^{3/2}$ for x=0.13 Ca doping.

**Fig.5**. Magnetoresistivity ratio $\Delta\rho/\rho(0)$ for Ca doping at x=0.02 as a function of B applied within the basal plane.

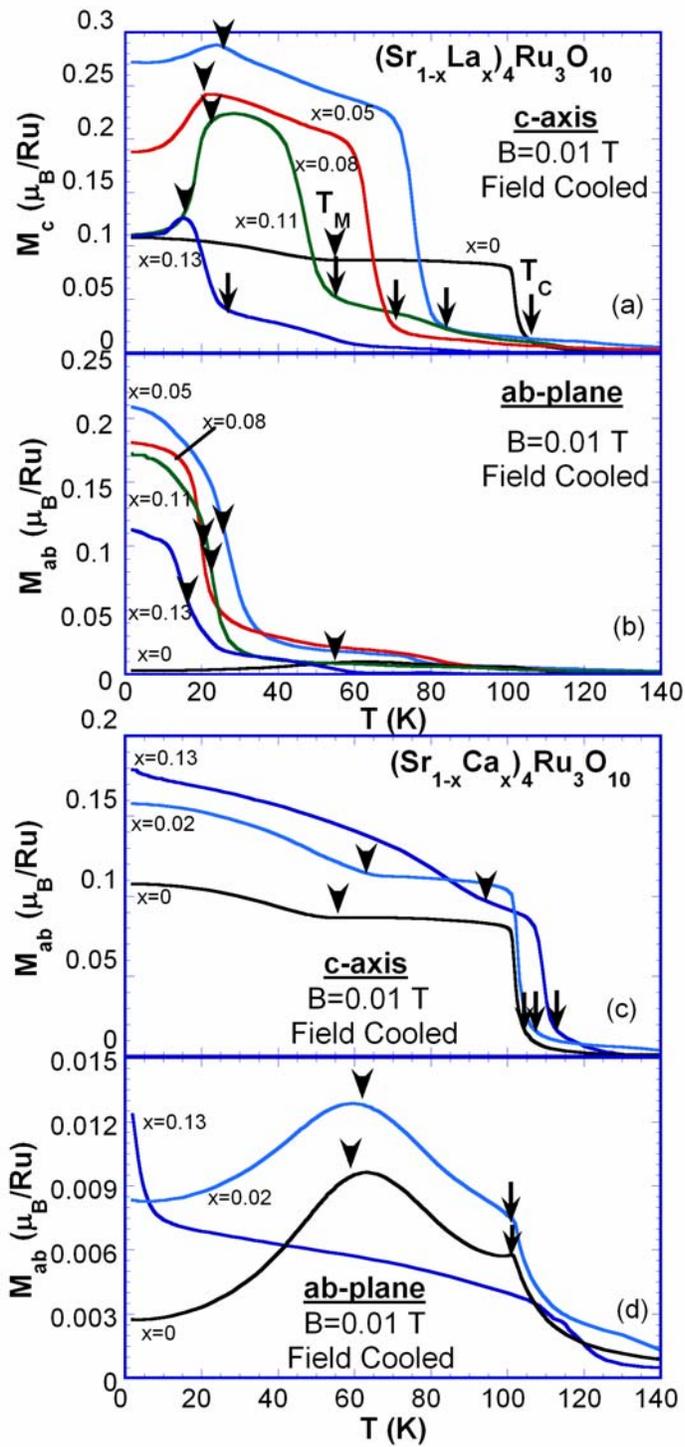

Fig. 1

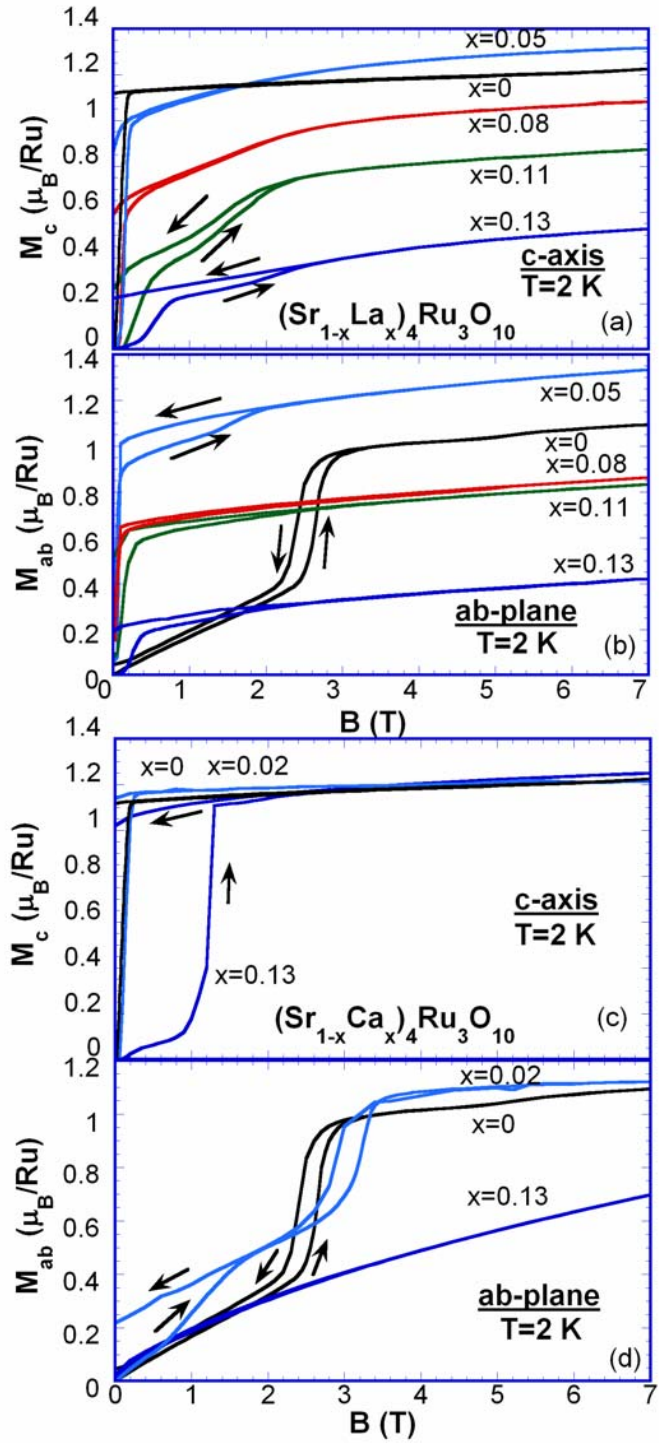

Fig. 2

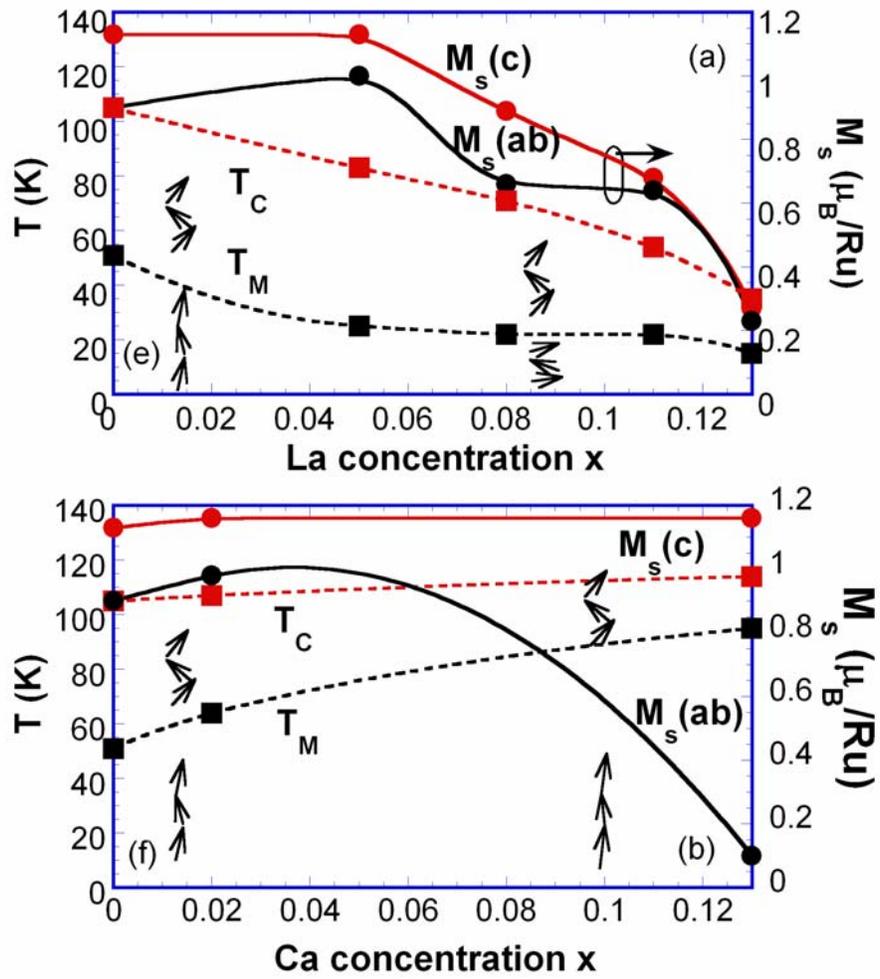

Fig. 3

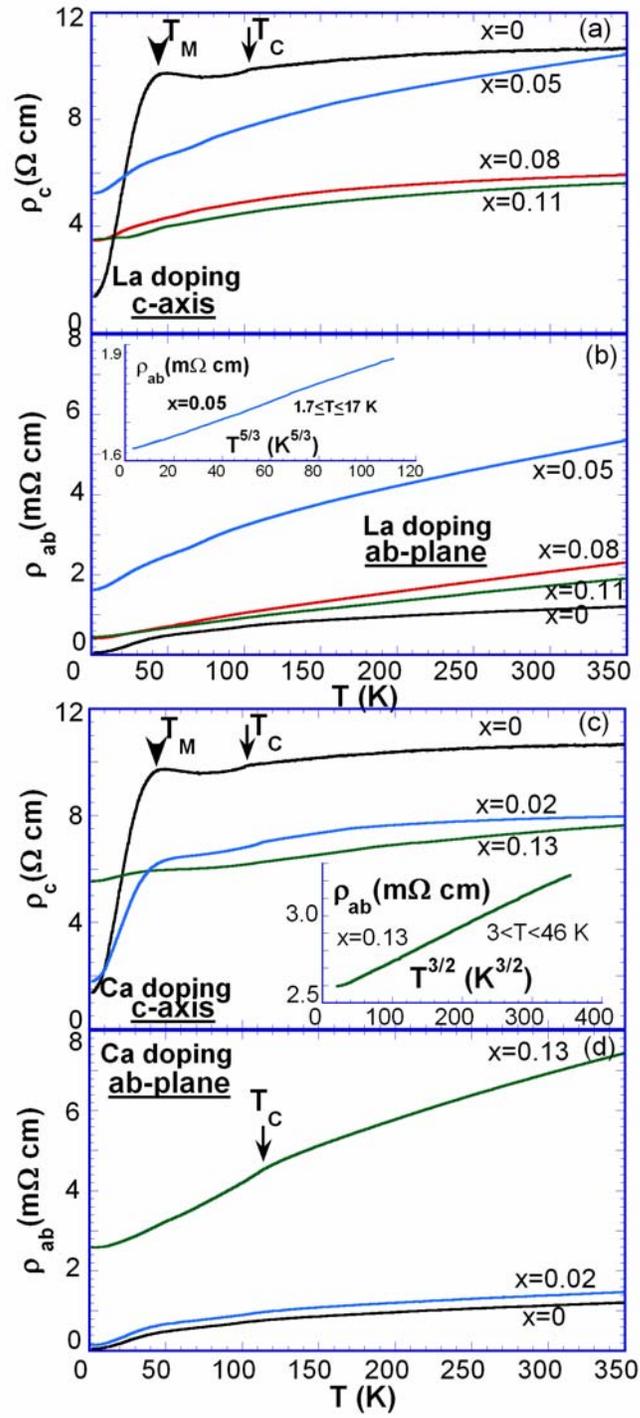

Fig. 4

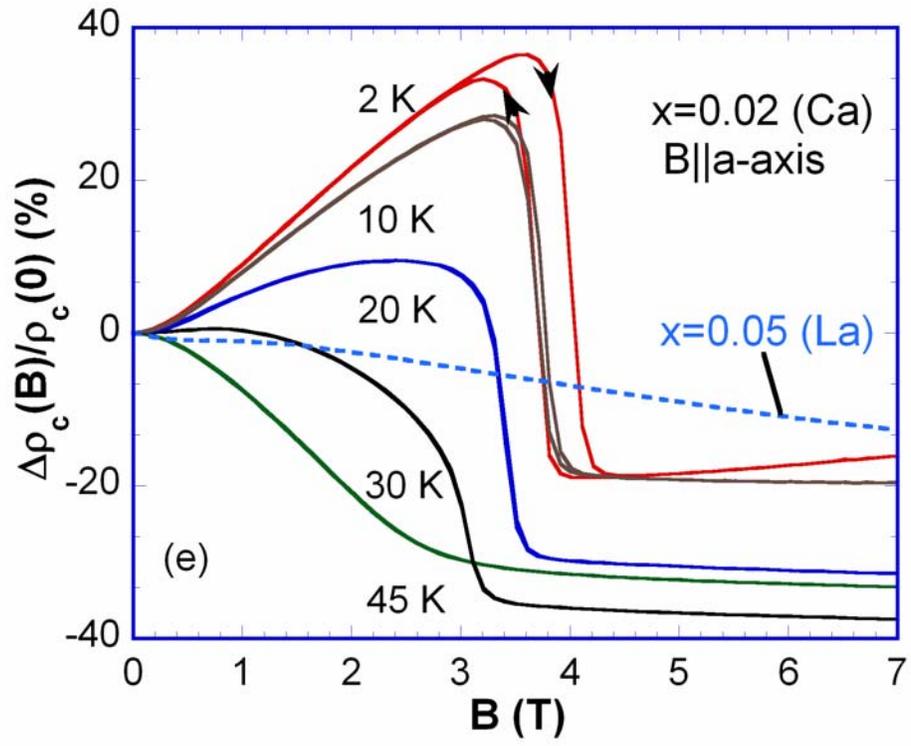

Fig. 5